\documentclass[11pt]{article}

\usepackage{amsfonts}
\usepackage{amssymb}
\usepackage{amsthm}

\def\GG{{\rm I}\!\Gamma}

\begin{document}
\begin{flushright}
{\bf IFUM 641/FT} \\
{\bf MPI/PhT-99-20}
\end{flushright}
\vskip 2 truecm
{
\Large
\bf
\centerline{Direct Algebraic Restoration of Slavnov-Taylor Identities}
\centerline{in the Abelian Higgs-Kibble Model}
\normalsize
\rm
\vspace{0.7cm} 
}

\begin{center}
{\Large Ruggero Ferrari$^a$\footnote{ruggero.ferrari@mi.infn.it}, 
Pietro Antonio Grassi$^b$\footnote{pgrassi@mppmu.mpg.de}, and 
Andrea Quadri$^a$\footnote{quadri@sunmite.mi.infn.it}}
\\
\vskip 0.5 truecm
$^a$ Dipartimento di Fisica, Universit\`a di Milano
via Celoria 16, 20133 Milano, Italy \\
and  INFN, Sezione di Milano \\
$^b$Max-Planck-Institut f\"ur Physik
(Werner-Heisenberg-Institut)
F\"ohringer Ring 6,\\ 80805 Munich, Germany

\end{center}
\setlength {\baselineskip}{0.2in}
\vskip 0.6 cm
\normalsize
\bf
\centerline{Abstract}
\rm
\small
\begin{quotation}
A purely algebraic method is devised in order to recover Slavnov-Taylor
identities (STI), broken by intermediate renormalization. The
counterterms are evaluated order by order in terms of finite
amplitudes computed at zero external momenta.
The evaluation of the breaking terms of the STI is avoided and their 
validity  is imposed directly on the vertex functional.  
The method is applied to the abelian Higgs-Kibble model.
An explicit mass term for the gauge field is introduced,
in order to check the relevance of nilpotency.
We show that,
since there are no anomalies, the imposition of the STI
turns out to be equivalent to the solution of a linear problem.
The presence of ST invariants implies that there are many possible
solutions, corresponding to different normalization conditions.
Moreover, we find more equations than unknowns (over-determined
problem). This leads us to the consideration of consistency
conditions, that must be obeyed if the restoration of STI
is possible.
\end{quotation}
\vfill\eject

\normalsize


\section{Introduction}

In quantum field theory, many essential physical requirements
can be expressed by means of Ward identities, valid for
the vertex functional  $\Gamma$.
Translating the invariance of the theory
under a certain symmetry in a functional form, Ward identities impose
several constraints on the possible structure of $\Gamma$. In
perturbation theory, renormalization schemes must fulfill these 
constraints (we speak in this case of an 
{\em invariant} 
action $\Gamma$), in order to get a finite, physically relevant
theory.
But very often regularization and subtraction procedures
are unable to respect all the symmetries of the theory,
producing Green functions that (although finite) are not 
invariant and hence not correct.

In the absence of anomalies \cite{PS}, one can recover 
exact Green functions, generated
by the correct effective action 
(henceforth denoted by $\GG$), by introducing
counterterms, designed to compensate the breaking of Ward
identities occurring for $\Gamma$.

In a preceding paper \cite{FG} a strategy was proposed to construct
these counterterms in gauge theories for the case of BRST symmetry
\cite{BRST,becHK} and
related Ward identities (STI).
The strategy was applied to the abelian Higgs-Kibble (HK) model
\cite{becHK,higgs}
and was based on the knowledge of the breaking terms
of the STI.
The STI for the HK model are (appendix A)
\begin{eqnarray}
S(\GG)= \int d^4 x \left[\partial^\mu c \, {\delta \GG
\over \delta {A^\mu}}
+\left ( \partial^\mu A_\mu + {{ev}\over{\alpha}}\phi_2 \right )
{\delta \GG \over \delta {\bar c}} \right] + \left(\GG,\GG\right)
\label{i1}
\end{eqnarray}

The parenthesis is defined as
\begin{eqnarray}
\left(X,Y \right) = 
\int d^4x \left [ {{\delta X}\over{\delta J_1}}{{\delta
Y}\over{\delta\phi_1}} 
+ {{\delta X}\over{\delta J_2}}{{\delta
Y}\over{\delta\phi_2}}  
-{{\delta X}\over{\delta \psi}}{{\delta
Y}\over{\delta{\bar\eta}}} 
+{{\delta X}\over{\delta {\bar\psi}}}{{\delta
Y}\over{\delta\eta}}
\right ]
\end{eqnarray}

$J_1, J_2, \eta, \bar \eta$ are external sources coupled to
non-linear BRST variations of the fields $\phi_1, \phi_2, \bar \psi,
\psi$.
It is convenient to introduce the linearized ST operator
\footnote{Notice that $\GG^{(0)}=\Gamma^{(0)}$ (the classical
action) is a known quantity.}:
\begin{eqnarray}
S_0(\GG) \equiv 
\int d^4 x \left[\partial^\mu c \, {\delta \GG \over \delta {A^\mu}}
+\left ( \partial^\mu A_\mu + {{ev}\over{\alpha}}\phi_2 \right )
{\delta \GG \over \delta {\bar c}} \right] +
\left(\GG^{(0)},\GG\right)
+ \left(\GG,\GG^{(0)}\right).
\end{eqnarray}
In order to get the counterterms at the $n$-th order
of perturbation theory, one can construct a 
functional $\Xi^{(n)}$ whose image under
the linearized Slavnov-Taylor operator is a quantity
expressed by means of  the lower-order symmetric Green functions
$\GG^{(j)}$, $j <n$, and  the  renormalized Green functions $\Gamma^{(n)}$ 
at the $n$-th order.

In this paper we perform the construction of $\GG$ by
the direct imposition of the STI
in the Algebraic Renormalization framework. Thus we
avoid the explicit calculation of the breaking terms.
The resulting recursive construction turns out to be
much simpler.
The problem consists in the solution of a linear set
of equations. In general, the number of equations exceeds
the number of unknowns (over-determined problem) and moreover
many solutions are possible, due to the existence of ST invariants.
The last property allows the imposition of the normalization 
conditions. The former property yields a set of consistency
conditions.
Their origin is  ascribable to the Quantum Action Principle (QAP)
\cite{PS,QAP} and to
the STI. The r\^ole of the nilpotency of $S_0$ is not very
clear in the present model. To illustrate this peculiar aspect of
the abelian model we introduce  an explicit mass term.
Such a theory is BRST invariant and moreover STI are valid due to the 
absence of any
anomaly. However the nilpotency of $S_0$ is broken by the mass term.
We verify that the consistency conditions are still valid, 
although modified by
the presence of the explicit mass term. Thus, in some sense, one is
tempted to conclude that the consistency conditions are valid even if
$S_0$ is not nilpotent.  The last conclusion is stated in a
conditional form, because the model under consideration violates
physical unitarity  \cite{unitarity}, and therefore the conclusion could not
be pertinent for physically relevant models. 

Finally we stress that the present procedure of the direct 
imposition of the STI turns out to be very efficient for 
deriving an algorithm which is implementable on computer. 
Therefore the aim of this paper is to provide a
preparatory work which can be translated to the Standard Model 
and its extensions. There more involved functional identities 
and the big amount of possible candidates for breaking terms
and counterterms require an analysis by means of symbolic
manipulation.
\section{Direct algebraic imposition of STI} \label{metodo}

The QAP implies that the breaking terms which spoil the
symmetries are,  at the first non-trivial order, local and compatible
with the power-counting. Therefore they can be removed, 
in absence of anomalies, by non-invariant counterterms.
Since we are concerned only in the construction
of the counterterms we can replace $\GG$ with
its effective part (i.e. the Taylor expansion of any
amplitude in the independent external momenta around zero
\footnote{in the absence of IR problems, as it is in the HK model.}
).
In this way 
 we associate to $\GG^{(n)}$ a formal series 
given by an infinite sum of local
Lorentz-invariant functionals. 
 Without possibility of confusion we can denote the series by
$\GG^{(n)}$ itself 
\begin{equation}
\GG^{(n)} = \sum_j \int d^4x \, m_j^{(n)} {\cal M}_j(x)
\label{e1}
\end{equation}

The Lorentz-  scalar monomials ${\cal M}_j(x)$ in the fields
and external sources (and their derivatives) have to comply with all
unbroken symmetries of the theory.
In the HK model, we require them to be $C$-even and
with zero FP-charge. We choose them linearly independent.
Notice that the expansion of $\GG$ in
eq. (\ref{e1}) may contain monomials  ${\cal M}_j$ with 
arbitrary positive dimension.

The coefficients $m^{(n)}_j$ are uniquely determined
once  the basis $\{ {\cal M}_j\}_{j \in {\bf N}}$ 
has been fixed.
The vector space spanned by $\{ {\cal M}_j \}_{j \in {\bf N}}$ 
is denoted
by ${\cal V}$.
\par
We have to impose recursively on $\GG$ the validity of the STI
\begin{equation}
S(\GG)=0
\label{e2}
\end{equation}
We expand the above equation in powers of $\hbar$. The contribution
to the $n$-th order is
\begin{equation}
\left [ S(\GG) \right ]^{(n)} = 
S_0(\GG^{(n)}) + \sum_{i=1}^{n-1} (\GG^{(n-i)}, \GG^{(i)}) = 0
\label{e3}
\end{equation}
The second term is given by $\GG^{(j)}$ with $j<n$. We assume
that $\GG^{(j)}$, $j<n$ satisfy STI.
The unknown quantities are the action-like parts of $\GG^{(n)}$,
(i.e. monomials with dimensions less or equal to four with the
correct symmetry properties)
which
we denote by $\Xi^{(n)}$.
$\Gamma^{(n)}$ is constructed by using the counterterms
$\Xi^{(j)}$ with $j<n$ and it is finite.
In the expansion of eq.(\ref{e1}), $\Xi^{(n)}$ is  
given in terms of
${\cal M}_k(x)$
with ${\rm dim} \ {\cal M}_k(x) \leq 4$ and their
coefficients are denoted by $\xi_k^{(n)}$. 
In the same way, the superficially convergent part of 
$\Gamma^{(n)}$ is given in terms of
${\cal M}_k(x)$ with ${\rm dim} \ {\cal M}_k(x) >4$ and their
coefficients are denoted by $\gamma_k^{(n)}$.
We maintain the notation $m_j^{(n)}$ to indicate collectively 
both $\xi_j^{(n)}$ and $\gamma_j^{(n)}$.
In the HK model, 
$\left [ S(\GG) \right ] ^{(n)}$ is an element 
of a  vector space $\cal W$ spanned by all possible 
linearly independent Lorentz-invariant,
$C$-even monomials in the fields and external sources, with
FP charge $+1$. 
\par
We choose a basis $\{ {\cal N}_i(x) \}_{i=1,2,3,\dots}$ for
${\cal W}$. 
We insert decomposition (\ref{e1}) in eq. (\ref{e3}):
{\small
\begin{eqnarray}
 && \hspace{-1.5cm} \left [ S(\GG) \right ]^{(n)} = 
\sum_j m_j^{(n)} 
S_0 \left ( \int d^4x \, {\cal M}_j \right ) +
\sum_{i=1}^{n-1} \sum_{jj'}  
m_j^{(i)} m_{j'}^{(n-i)}
\left ( \int d^4x \, {\cal M}_j(x), 
\int d^4x' \, {\cal M}_{j'}(x') \right ) 
\label{e4}
\end{eqnarray}
}
There are coefficients $a^j_r$,
$b^{jj'}_{kr}$ (uniquely fixed by the choice of
${\cal N}_i(x)$ 
and by the ST transformations) such that
\begin{eqnarray}
&& S_0 \,  \Big ( \int d^4x \, {\cal M}_j  \Big ) = 
\int d^4x \, \sum_r a_r^j {\cal N}_r(x) \nonumber \\
&& \Big ( \int d^4x \, {\cal M}_j(x), \int d^4x' \, {\cal M}_{j'}(x')
\Big )  =
\int d^4x \, \sum_r b^{jj'}_{r} {\cal N}_r(x) 
\label{e5}
\end{eqnarray}
Then eq. (\ref{e4}) becomes 
\begin{equation}
\sum_j a^j_r m_j^{(n)} + \sum_{i=1}^{n-1}
\sum_{jj'} m_j^{(i)} m_{j'}^{(n-i)} b^{jj'}_{r}=0
\quad r=0,1,2,\dots
\label{e7}
\end{equation}
For $r$ such that $\hbox{dim } {\cal N}_r(x) >5$,
eq.~(\ref{e7}) is an identity by the virtue of the QAP.
In the absence of anomalies
(as it is in the HK model), one can solve eq.~(\ref{e7})
expressing coefficients $\xi^{(n)}_j$ in terms of the coefficients
$\gamma^{(n)}_j$ and $m_j^{(l)}$, $l<n$.
That is, we construct $\Xi^{(n)}$ from  
the superficially convergent  part of $\Gamma^{(n)}$, and from
lower order contributions $\GG^{(l)}$, $l <n$.

If we can reach this goal, then STI can be restored.
Notice that at every step one does not need to consider coefficients
$m_j^{(n)}$ whose associated monomials have dimension $>6$ (for the
HK model): they will never contribute to eqs. (\ref{e7})
for $r$ such that ${\rm dim} \, {\cal N}_r(x) \leq 5$.
This allows the recursive algebraic construction of $\GG$. However 
the procedure requires the fixing of the normalization conditions
(associated to the existence of ST invariants) and the use of
the consistency conditions. Both items have been discussed at length
in Ref. \cite{FG}. The normalization conditions are used in the
solution of the linear problem given by eqs. (\ref{e5}) and
(\ref{e7}), in order to simplify the algebra. The evaluation of
the physical S-matrix elements requires an analysis of the
two-point functions; in particular one must evaluate 
the location of the poles and 
their residua. The present approach makes use of the normalization
conditions in the same manner. However the consistency conditions
show up in a different way. In Ref. \cite{FG} the action
counter-terms $\Xi^{(n)}$ are obtained by solving the equation
\begin{equation}
S_0(\Xi^{(n)}) = -\Psi^{(n)}
\label{e7.1}
\end{equation}
where $\Psi^{(n)}$ is given in terms of finite amplitudes. A solution
exists only if 
\begin{equation}
S_0(\Psi^{(n)}) = 0
\label{e7.2}
\end{equation}
and this provides the consistency conditions (the use of the ghost
equation might be necessary if nilpotency of the BRST is only
on-shell). In the present approach the counter-terms are evaluated
by imposing STI directly on the effective vertex functional
and one doesn't need to evaluate $\Psi^{(n)}$. Moreover the method
makes no use of the nilpotency of the BRST. Therefore the consistency
conditions in the present approach must be the consequence of a
more general property than (\ref{e7.2}). In the next section the use of the
normalization conditions is briefly recalled. In order to illustrate
the problem of the consistency conditions we consider the HK model
with an explicit mass term $M$. The model is BRST symmetric but
nilpotency is broken by the explicit mass term. Furthermore it has
an unpleasant feature: physical unitarity is violated (as it can
be checked by an explicit calculation). However this fact is not
relevant in our discussion.

\section{The massive HK model}

We study the HK massive model, whose classical action is given in
Appendix A.
$\GG$ satisfies 
the ghost equation
\begin{equation}
{{\delta\GG}\over{\delta{\bar c}}} = (\alpha \square + M^2) c 
+ev{{\delta\GG}\over {\delta{J_2}}}
\label{e8}
\end{equation}

Eq.(\ref{e8}) implies
\begin{eqnarray}
& & 
ev \xi_{J_{2}c}^{(n)} = \xi_{\bar c c}^{(n)},~~~~ 
ev \xi_{J_{2}c \phi_{1}}^{(n)} = \xi_{\bar c c \phi_{1}}^{(n)},~~~~
ev \gamma_{J_{2} c \phi_{1}^{2}}^{(n)} = \xi_{\bar c c
\phi_{1}^{2}}^{(n)},~~~~ 
\nonumber \\
& & 
ev \gamma_{J_{2} c \phi_{2}^{2}}^{(n)} = \xi_{\bar c c
\phi_{2}^{2}}^{(n)},~~~~ 
ev \gamma_{J_{2} c A^{2}_{\mu}}^{(n)} = \xi_{\bar c c
A^{2}_{\mu}}^{(n)},~~~~ 
ev \gamma_{J_{2} \square c}^{(n)} = \xi^{(n)}_{\bar c \square c}
\label{e9}
\end{eqnarray}
so it fixes the following counterterms to all orders $n\geq 1$
\footnote{If not explicitly shown, the order of counterterms is
understood to be $n$.}
\begin{equation}
\xi_{{\bar c} \square c}, \xi_{{\bar c} c \phi_1^2}, \xi_{{\bar c} c
\phi_2^2}, \xi_{{\bar c} c A_\mu^2},
\label{e10}
\end{equation}
while $\xi_{\bar c c}, \xi_{\bar c c \phi_1}$ depend on the
normalization conditions imposed  on external sources
counterterms $\xi_{J_2c}, \xi_{J_2 c \phi_1}$.
We notice that the solution of eq.(\ref{e7}) is not unique. Consider
eq. (\ref{e3}) and suppose that $\Xi^{(n)}$ is a solution. Then, for
arbitrary c-numbers $v_j$, $\Xi^{(n)}+\sum_j v_j {\cal I}_j$
is again a solution, provided that $S_0[{\cal I}_j]=0$.
As in \cite{FG} we look for action-like $S_0$-invariants preserving
unbroken symmetries of the model. We find $11$ linearly independent
$S_0$-invariants ${\cal I}_1 - {\cal I}_{11}$, listed in Appendix B.
For the HK massive model, after taking into account
the ghost equation (\ref{e8}) the $S_0$ operator becomes
($n\geq 1$)
{\small
\begin{eqnarray}
{\hat S}_0(\GG^{(n)})&\equiv&
\int d^4x \left\{ \partial_\mu c{{\delta\GG}\over{\delta{A_\mu}}}^{(n)}
-ec\phi_2{{\delta\GG}\over{\delta{\phi_1}}}^{(n)}
+ ec(\phi_1+v){{\delta\GG}\over{\delta{\phi_2}}}^{(n)}
+ i {e\over 2} c{\bar\psi} \gamma_5
{{\delta\GG}\over{\delta{\bar \psi}}}^{(n)}
- i{e\over 2}
{{\delta\GG}\over{\delta{\psi}}}^{(n)}
\gamma_5\psi c 
\right. \nonumber \\
&+& \left. 
{{\delta\GG}\over{\delta{\phi_1}}}^{(0)}
{{\delta\GG}\over{\delta{J_1}}}^{(n)}
+ \left[
{{\delta\GG}\over{\delta{\phi_2}}}^{(0)}
+ev(\partial^\mu A_\mu +
{{ev}\over\alpha}\phi_2)\right]
{{\delta\GG}\over{\delta{J_2}}}^{(n)}
-
{{\delta\GG}\over{\delta{\psi}}}^{(0)}
{{\delta\GG}\over{\delta{\bar \eta}}}^{(n)}
+
{{\delta\GG}\over{\delta{\bar \psi}}}^{(0)}
{{\delta\GG}\over{\delta{\eta}}}^{(n)} \right\}
\nonumber \\
\label{e11}
\end{eqnarray}
}
The transformation
\begin{equation}
\Xi^{(n)} \rightarrow \Xi^{(n)}+\sum_j v_j {\cal I}_j
\label{e12}
\end{equation}
must be compatible with eq. (\ref{e8}).
This requirement entails the constraints
\begin{eqnarray}
&&v_7=0
\nonumber\\
&&
v_8+[(ev)^2+M^2] v_9 + e^2v ~v_{11} =0.
\label{e12.p}
\end{eqnarray}
All other invariants ${\cal I}_j$ can be used to put to zero some of
the
counterterms in $\Xi$, to all orders in perturbation theory.
Invariants ${\cal I}_9-{\cal I}_{11}$ are used to set to zero the
bosonic
external sources counterterms
\begin{equation}
\xi_{J_2 c \phi_1}, \xi_{J_2c}, \xi_{J_1c}.
\label{e13}
\end{equation}
On the contrary, it is not possible to put to zero fermionic 
external sources counterterms. Then one gets contributions
to $\Xi^{(n)}$ from the fermionic mass term.               
The remaining ${\cal I}_j$ are used to put to zero the following
counterterms:
\begin{eqnarray}
&& \xi_{\phi_1}, \xi_{\phi_2^2\phi_1}, \xi_{A_\mu^2 \phi_1},
\xi_{F_{\mu \nu}^2},\nonumber \\
&& \xi_{\bar \psi \gamma_\mu \gamma_5 \psi A^\mu}, \xi_{i \bar \psi
\gamma_5 \psi \phi_2}.
\label{e14}
\end{eqnarray}
$\Xi^{(n)}$ is then completely determined from the set of linear
equations
(\ref{e7}), once we adopt the normalization conditions fixed in
eqs.(\ref{e13})
and (\ref{e14}). In particular, conditions in eq.(\ref{e14}) are
obtained
according to the introduction of a hierarchy in the counterterms,
designed
to get maximal decoupling among equations (\ref{e7}). 
Consider two action-like monomials ${\cal M}_j(x)$ and ${\cal
M}_{j'}(x)$
such that their $S_0$-images do not contain common terms; then their
coefficients $\xi_j$ and $\xi_{j'}$ can never appear in the same
equation in (\ref{e7}): they do not couple.
We decompose the counterterms $\Xi^{(n)}$ in disjoint sectors, 
thanks to normalization conditions (\ref{e13}) and (\ref{e14}).
Next we find it convenient to solve eqs.(\ref{e7}) for amplitudes
with lower number of external legs and higher derivatives in the
external
momenta. This reduces the number of graphs that must be evaluated
(at the cost of a higher number of derivatives).
Notice that the analysis of counterterms sectors is indeed equivalent
to the construction of coefficients $a_r^j$ in eq.(\ref{e5}), by
taking into
account the normalization conditions in eqs. (\ref{e13}), (\ref{e14}).
Since the difference in $S_0$ between the massive and massless HK
models
only involves terms depending on $J_1$, $J_2$, the same counterterms
sectors can be used in the massive case as well as in the massless
one.
We recover the conventional HK model by taking the limit 
$M \rightarrow 0$ in the expression of $\Xi^{(n)}$. This limit is
smooth.
\section{Consistency conditions} \label{cc}
In the Higgs-Kibble boson-gauge sector there are more equations than
unknowns (over-determined problem).
We then find a set of 
consistency conditions associated to the linear problem. For the massive case 
they are: 
\begin{eqnarray}
&&
\hat m_1^2 \left [ ev \left ( \gamma_{\partial_\mu J_1 c \partial^\mu \phi_2} - 
\gamma_{J_1 c \square \phi_2} \right )
+ e \gamma_{\square J_2 c} + \left ( \gamma_{\partial_\mu J_1 c
A^\mu} - \gamma_{J_1c \partial_\mu A^\mu} \right ) \right ] \nonumber \\
&&
\qquad +ev {M^2 \over \alpha} 
\left ( \gamma_{\partial_\mu J_2 c \partial^\mu \phi_1}
-\gamma_{\square J_2 c \phi_1} \right ) = 0 
\label{cc1} \\
\nonumber \\
&& 
6 \lambda v \Big (  \gamma_{J_1 c \partial_\mu A^\mu}
                      -\gamma_{\partial_\mu J_1 c A^\mu} \Big )
  -\hat m_1^2 \gamma_{J_1 c A^\mu \partial_\mu \phi_1}
  -\hat m_1^2 \gamma_{\partial_\mu J_1 c A^\mu \phi_1}
  +2 \hat m_1^2 \gamma_{J_1 c \partial_\mu A^\mu \phi_1}
\nonumber \\
&& -e\hat m_1^2 \gamma_{\partial_\mu J_1 c \partial^\mu \phi_2}
   -2e^2v \gamma_{\partial_\mu \phi_2 \partial^\mu \phi_2 \phi_1}
   -4\lambda e v \gamma_{\square J_2c}
   -2ev \gamma_{J_2 c \phi_1^2}	-e{M^2 \over \alpha} \gamma_{\partial_\mu J_2 c \partial^\mu \phi_1}
\nonumber \\
&& +2ev \Big ( \gamma_{\phi_2 \partial_\mu \phi_1^2 A^\mu}
              -\gamma_{\phi_2 \phi_1^2 \partial_\mu A^\mu}
        \Big )
   + 6 \sum_{j=1}^{n-1} \Big ( \gamma^{(j)}_{\partial^\mu J_1 c
A_\mu}
                             - \gamma^{(j)}_{J_1c \partial^\mu A_\mu}
\Big )
                        \xi_{\phi_1^3}^{(n-j)} =0
\label{cc2} \\
\nonumber \\
&&
-2\hat m_1^{2} \gamma_{J_{1}c \phi_{2}A^{2}} -4v \lambda \gamma_{J_{2}cA^{2}} +
2e^2v \gamma_{J_{1}c \phi_{2}\phi_{1}}
+4e\lambda v \gamma_{\partial_\mu J_1 c A^\mu} 
-e^2v \gamma_{\phi_2\phi_1 \partial_\mu \phi_1 A^\mu} \nonumber \\
&&
+e\hat m_1^2 \gamma_{J_1cA^\mu \partial_\mu \phi_1} 
+e\hat m_1^2 \gamma_{\partial_\mu J_1 c A^\mu \phi_1} 
+2\sum_{j=1}^{n-1} \gamma^{(j)}_{J_1c\phi_2\phi_1} \xi^{(n-j)}_{\phi_1 A^2}
-6e \sum_{j=1}^{n-1} \gamma^{(j)}_{\partial^\mu J_1 c A_\mu} \xi^{(n-j)}_{\phi_1^3} =0
\label{cc3} \\
\nonumber \\
&&
e{M^2 \over \alpha} \gamma_{\partial_\mu J_2 c \partial^\mu \phi_1}
+2e^2v \gamma_{\partial_\mu \phi_2 \partial^\mu \phi_2 \phi_1}
+2 \lambda v (\gamma_{J_1c \partial_\mu A^\mu} -
\gamma_{\partial_\mu J_1 c A^\mu}) \nonumber \\
&&
\qquad 
+e \hat m_1^2 \gamma_{\partial_\mu J_1 c \partial^\mu \phi_2}
-6ev \gamma_{\phi_2^3 \partial_\mu A^\mu}
-2ev \gamma_{J_2c\phi_2^2} +4e\lambda v \gamma_{\square J_2 c} = 0
\label{cc4} \\
\nonumber \\
&& ev \Big [ \gamma_{\phi_{2}A_{\mu}\partial^{\mu}A^{2}}
+e \gamma_{\partial_\mu J_{1}c \partial_{\mu}A^{\mu}} \Big ]
+\sum_{j=1}^{n-1} \gamma_{\partial_\mu J_{1}c A^{\mu}}^{(j)} 
\xi_{\phi_{1}A^{2}}^{(n-j)} = \nonumber \\
&& \qquad ev \Big [ \gamma_{\phi_{2}\partial^{\mu}A_{\mu}A^{2}} +
e \gamma_{J_{1}c\partial_\mu A^{\mu}} + \gamma_{J_2 c A^2} \Big ] 
+ \sum_{j=1}^{n-1} \gamma^{(j)}_{J_{1}c\partial^\mu A_{\mu}} 
\xi_{\phi_{1}A^{2}}^{(n-j)}.
\label{cc5}
\end{eqnarray}
In the Fermi sector
\begin{eqnarray}
ev \gamma_{\phi_2{\bar \psi} \gamma^{\mu} \psi A_\mu  } + 
G\, v\Big( \gamma_{c {\bar \psi } \gamma^{\mu}  \eta A_\mu}
- 
\gamma_{c {\bar \eta } \gamma^{\mu}  \psi A_\mu}\Big)+
\sum_{j=1,n-1}
\xi_{{\bar \psi} \psi}^{(j)}\Big( \gamma^{(n-j)}_{c {\bar \psi }
\gamma^{\mu}\eta A_\mu}
-  
\gamma^{(n-j)}_{c {\bar \eta } \gamma^{\mu}\psi A_\mu}
\Big) =0 .
\label{fermi.33}
\end{eqnarray}


$\hat m_1^2$ denotes the quantity $2 \lambda v^2 + {M^2 \over
\alpha}$. 
The above consistency conditions can be obtained from a quite
general equation. 
Let us introduce the linearized version $S_{F}$ of the ST operator 
$S(F)$ defined in eq.(\ref{i1}) for a generic functional $F$:
\begin{equation}
S_{F}(\cdot) = 
\int d^4 x \left[\partial^\mu c \, {\delta (\cdot) 
\over \delta {A^\mu}}
+\left ( \partial^\mu A_\mu + {{ev}\over{\alpha}}\phi_2 \right )
{\delta  (\cdot) \over \delta {\bar c}} \right] + ({F}, \cdot) + (\cdot, {F}).
\label{ncc1}
\end{equation}
By straightforward algebra one obtains the identity
\begin{equation}
S_{F} S({F}) = \int d^4 x \left(\square c +{{ev}\over{\alpha}}
{\delta {F} \over \delta J_2}\right)
{\delta {F} \over \delta \bar c}.
\label{ncc2.1}
\end{equation}
If $F$ obeys the ghost equation (\ref{e8}), then
\begin{equation}
S_{F} S({F}) = -{M^2 \over \alpha}  \int d^4 x ~ c
{\delta {F} \over \delta \bar c}.
\label{ncc2}
\end{equation}
Suppose that we have restored the STI up to the $(n-1)$-th order
in perturbation theory, i.e. $S(\GG)^{(j)}=0$ for
$j=0,1,\dots,n-1$. 
At the $n$-th order of perturbation theory
equation (\ref{ncc2}) implies
\begin{equation}
S_0 [S(\Gamma)^{(n)}]= -{M^2 \over \alpha} \int d^4 x~ c
{\delta {\Gamma} \over \delta \bar c}^{(n)}
\label{ncc3}
\end{equation}
Now we prove that in the abelian HK model the r.h.s. is zero. 
Let us consider the breaking
terms of the STI at the order $n$
\begin{equation}
\Delta\Gamma^{(n)}\equiv S(\Gamma)^{(n)}.
\label{ncc2.3}
\end{equation}
From the QAP we know that $\Delta\Gamma^{(n)}$ is a local
functional with dimension less or equal five and FP-charge
equal one. Then by construction
\begin{equation}
 -{M^2 \over \alpha}\int d^4x ~ c
{\delta {\Gamma} \over \delta \bar c}^{(n)}=
S_0 [S(\Gamma)^{(n)}]=S_0 [\Delta\Gamma^{(n)}]
\label{ncc3.4}
\end{equation}
i.e $c{{\delta\Gamma}\over{\delta{\bar c}}}^{(n)} $ is a local functional and 
has dimension less or equal four
and FP-charge equal two. There are no terms with these properties
($\int d^4x c\square c =0$,$\int d^4x c c =0$,  $\int d^4x
cA^\mu\partial_\mu c$
is not allowed by C-conjugation, etc.).
This implies that 
\begin{equation}
S_0 [S(\Gamma)^{(n)}]= 0.
\label{ncc3p}
\end{equation}
On account of the results obtained in sect. \ref{metodo},
$S(\Gamma)^{(n)}$ can be expanded on monomials ${\cal N}_r(x)$
(see eqs.(\ref{e4}),(\ref{e5})) with dimension $\leq 5$, whose
coefficients are constructed according to eq.(\ref{e7}).
The imposition of eq.(\ref{ncc3p}) then yields a
set of consistency conditions 
equivalent to eqs.(\ref{cc1})-(\ref{cc5}).
\par
We illustrate this procedure for the consistency condition in
eq.(\ref{cc5}).
We denote by $n_{c \partial_\mu A^\mu A^2}$ the coefficient
of the monomial $c \partial_\mu A^\mu A^2$
in the expansion of $S(\Gamma)^{(n)}$, and by
$n_{c A^\mu \partial_\mu A^2}$ the analogous coefficient for
$c A^\mu \partial_\mu A^2$.
The coefficient of the monomial $A^2\square c c$ in 
$S_0[S(\Gamma)^{(n)}]$ turns out to be equal to 
$n_{c \partial_\mu A^\mu A^2} - n_{c A^\mu \partial_\mu A^2}$. 
This must be zero according to eq.(\ref{ncc3p}):
\begin{equation}
n_{c \partial_\mu A^\mu A^2} - n_{c A^\mu \partial_\mu A^2}=0
\label{ncc4p}
\end{equation}
The method of sect.\ref{metodo} allows the explicit evaluation of
$n_{c \partial_\mu A^\mu A^2}$ and  $n_{c A^\mu \partial_\mu A^2}$.
According to eq.(\ref{e7}), we get
\begin{eqnarray}
n_{c \partial_\mu A^\mu A^2} & = &
ev \Big [ \gamma_{\phi_{2}\partial^{\mu}A_{\mu}A^{2}} +
e \gamma_{J_{1}c\partial_\mu A^{\mu}} + \gamma_{J_2 c A^2} \Big ] 
-4 \xi_{A^4}
+ \sum_{j=1}^{n-1} \gamma^{(j)}_{J_{1}c\partial^\mu A_{\mu}} 
\xi_{\phi_{1}A^{2}}^{(n-j)}
\label{nf1}
\end{eqnarray}
and
\begin{eqnarray}
n_{c A^\mu \partial_\mu A^2} & = &
 ev \Big [ \gamma_{\phi_{2}A_{\mu}\partial^{\mu}A^{2}}
+e \gamma_{\partial_\mu J_{1}c \partial_{\mu}A^{\mu}} \Big ]
-4 \xi_{A^4}
+\sum_{j=1}^{n-1} \gamma_{\partial_\mu J_{1}c A^{\mu}}^{(j)} 
\xi_{\phi_{1}A^{2}}^{(n-j)} 
\label{nf2}
\end{eqnarray}
Inserting eqs.(\ref{nf1}) and (\ref{nf2}) in
eq.(\ref{ncc4p}) we finally recover the consistency condition (\ref{cc5}).
\section{Conclusions}
In the present paper we propose a method for imposing STI
directly on the effective vertex functional (its formal
Taylor series expansion at zero external momenta), without going
through the explicit evaluation of the breaking terms. 
The method can be applied to any (not symmetrically) renormalized 
gauge theory where the
renormalization procedure (regularization and subtraction of
divergent parts) violates STI. 
The algorithm amounts
to find a basis of all action-like local Lorentz-invariant monomials
and their ST transforms. Preserved symmetries have to be imposed
on the allowed monomials (C-conjugation, FP charge neutrality, etc.). 
BRST sources have to be considered in the construction of the
monomials. The counter-terms are constructed order by order by solving
a linear problem where the input data are a set of finite zero-momenta
amplitudes of dimension five and six. The existence of a set
of ST invariants allows to fix an equivalent number of normalization
conditions. These are chosen in order to simplify the solution
and the evaluation of the finite amplitudes (normalization
conditions at zero momenta). There is a number of consistency
conditions associated to the linear problem. We suggest the use
of a general property of the STI ($S_0 [S(\Gamma)^{(n)}]=
0$ for the example considered here, i.e. an abelian gauge model) 
in order to find
the necessary consistency conditions that should be satisfied
by the finite amplitudes relevant for the linear problem. 
\section{Acknowledgment}
We acknowledge a partial financial support by MURST.
\appendix

\section{Classical action}
The classical action for the massive HK model is 
\begin{eqnarray}
&&
\Gamma^{(0)} = \int d^4x \Big [- {1\over 4}F_{\mu\nu}^2
+ {{e^2v^2}\over 2} A_\mu^2
\nonumber\\
&&
-{\alpha\over 2}\partial A^2 + \alpha {\bar c}\square c 
+ e^2v^2{\bar c}c + e^2 v {\bar c}c\phi_1
\nonumber\\
&&
+{1\over 2}((\partial_\mu\phi_1)^2 + (\partial_\mu\phi_2)^2)
-\lambda v^2 \phi_1^2 - 
{{e^2v^2}\over {2\alpha}} \phi_2^2
\nonumber\\
&&
+eA_\mu (\phi_2\partial^\mu \phi_1-\partial^\mu\phi_2\phi_1)
+ e^2v\phi_1 A^2 +{{e^2}\over 2}(\phi_1^2+\phi_2^2) A^2 
\nonumber\\
&&
- \lambda v \phi_1(\phi_1^2+\phi_2^2)
-{\lambda\over 4}(\phi_1^2+\phi_2^2)^2
\nonumber\\
&&
+ {\bar\psi}i\not \!\!\partial\psi +Gv{\bar\psi}\psi
+{e\over 2}{\bar\psi}\gamma_\mu\gamma_5\psi A^\mu
\nonumber\\
&&
+G{\bar\psi}\psi\phi_1
-iG{\bar\psi}\gamma_5\psi\phi_2
\nonumber\\
&&
+J_1 [-ec\phi_2] + J_2 ec(\phi_1+v)
+i{e\over 2}{\bar\eta}\gamma_5\psi c
+i{e\over 2}c{\bar\psi}\gamma_5 \eta
\nonumber \\
&& + {M^2 \over 2} A_\mu^2 + M^2 \bar c c 
- {M^2 \over 2 \alpha} (\phi_1^2 + \phi_2^2) \Big ] 
\end{eqnarray}

BRST transformations
\begin{eqnarray}\label{BRST}
&&
s A_\mu = \partial_\mu c, ~~~ 
s \phi_1 = -ec \phi_2, ~~~ 
s \phi_2 = ec (\phi_1 +v) 
\nonumber \\
&&
s \psi = -i {e \over 2} \gamma_5 \psi c, ~~~ 
s \bar \psi = i {e \over 2} c \bar \psi \gamma_5, ~~~ 
s \bar c = \partial A + {ev \over \alpha} \phi_2, ~~~ 
s c = 0
\end{eqnarray}

\section{$S_0$ invariants}
\begin{eqnarray}
&&
{\cal I}_1 = \int d^4x (\phi_1^2 + \phi_2^2
+2 v\phi_1)
\nonumber\\ 
&&
{\cal I}_2 = \int d^4x (\phi_1^4 + \phi_2^4
+ 2\phi_1^2\phi_2^2 + 4 v\phi_1^3 + 4 v \phi_1\phi_2^2
+4v^2\phi_1^2)
\nonumber\\ 
&&
{\cal I}_3 = \int d^4x |D_\mu \phi|^2
\nonumber\\ 
&&
{\cal I}_4 = \int d^4x (F_{\mu\nu})^2
\nonumber\\ 
&&
{\cal I}_5 = \int d^4x {\bar \psi}i\gamma_\mu {\cal D}^\mu \psi
\nonumber\\ 
&&
{\cal I}_6 = \int d^4x {\bar \psi}[(\phi_1+v) -
i\gamma_5\phi_2]\psi
\nonumber\\ 
&&
{\cal I}_7 = \int d^4x({1\over 2}{\cal F}^2 +
{\bar c}\delta_{\rm BRST}{\cal F})
\nonumber\\ 
&&
{\cal I}_{8} = \int d^4x ({1\over 2} A^2 +{\bar c}c + 
{v\over\alpha}\phi_1 ) 
\nonumber \\
&&
{\cal I}_{9}  = 
\int d^4x [A^\mu\Gamma^{(0)}_{A^\mu}+
c\Gamma^{(0)}_c
+\alpha ({\cal F}\partial^\mu A_\mu - {\bar c}\square c)
+ {M^2 \over 2\alpha} (\phi_1^2 + \phi_2^2)]
\nonumber\\ 
&&
{\cal I}_{10} = 
S_0 (\int d^4x J_1), ~~~~~~~
{\cal I}_{11} = S_0(\int d^4x J_1\phi_1) 
\end{eqnarray}
\begin{eqnarray*}
D_\mu = \partial_\mu -ie A_\mu, 
{\cal D}_\mu = \partial_\mu -i {e \over 2} \gamma_5 A_\mu
\end{eqnarray*}

\end{document}